\documentclass[sigconf]{acmart}
\usepackage{tabularx}
\usepackage{booktabs}   
\usepackage{tabularx}   
\usepackage{ragged2e}   
\usepackage{array}      
\usepackage{enumitem}
\usepackage{acmart-taps}
\newlist{designgoals}{enumerate}{1}
\setlist[designgoals]{
  label=\textbf{[DG\arabic*]},
  ref=DG\arabic*,
  leftmargin=*
}

\AtBeginDocument{%
  }

\usepackage{tabularx}
\usepackage{booktabs}
\usepackage{xcolor}     
\copyrightyear{2026}
\acmYear{2026}
\setcopyright{cc}
\setcctype{by}
\acmConference[CHI '26]{Proceedings of the 2026 CHI Conference on Human Factors in Computing Systems}{April 13--17, 2026}{Barcelona, Spain}
\acmBooktitle{Proceedings of the 2026 CHI Conference on Human Factors in Computing Systems (CHI '26), April 13--17, 2026, Barcelona, Spain}
\acmPrice{}
\acmDOI{10.1145/3772318.3791482}
\acmISBN{979-8-4007-2278-3/2026/04}

\begin{document}

\title{Gamifying Compassion: Mitigating Dialect Prejudice Through An AI-Driven Serious Game}

\author{Sicheng Lu}
\orcid{0009-0008-2976-4262}
\authornote{These authors contributed equally.}

\affiliation{%
  \institution{Xi'an Jiaotong-Liverpool University}
  \city{Suzhou}
  \state{Jiangsu}
  \country{China}
}
\email{Sicheng.Lu24@student.xjtlu.edu.cn}

\author{Erick Purwanto}
\orcid{0000-0001-6497-6721}
\authornotemark[1]
\authornote{Corresponding author.} 

\affiliation{%
  \institution{Xi'an Jiaotong-Liverpool University}
  \city{Suzhou}
  \state{Jiangsu}
  \country{China}
}
\email{Erick.Purwanto@xjtlu.edu.cn}

\author{Hong Liu}
\orcid{0000-0002-1154-4017}
\affiliation{%
  \institution{Xi'an Jiaotong-Liverpool University}
  \city{Suzhou}
  \state{Jiangsu}
  \country{China}
}
\email{Hong.Liu@xjtlu.edu.cn}

\author{Adel Chaouch-Orozco}
\orcid{0000-0003-0102-1754}
\affiliation{%
  \institution{City University of Hong Kong}
  \city{Hong Kong SAR}
  \country{China}
}
\email{adel.chaouchorozco@cityu.edu.hk}

\author{Aini Li}
\orcid{0000-0002-5190-8065}
\authornotemark[2]
\affiliation{%
  \institution{City University of Hong Kong}
  \city{Hong Kong SAR}
  \country{China}
}
\email{ainili@cityu.edu.hk}

\renewcommand{\shortauthors}{Lu et al.}

\begin{abstract}
  Dialect bias is pervasive yet often unconscious, normalized, or obscured by masking. Existing HCI interventions primarily audit disparities and propose reactive fixes. We present \textbf{CompassioMate}, a dialect-aware serious game that nurtures perspective-taking through AI-mediated play. Players listen to audio samples to identify regional dialects, engage in simulated social interactions involving dialect discrimination, and explore branching narratives that reveal how changes in wording or stance can influence the outcomes. In a three-week field study with 20 university students, participants reported feeling comfortable when observing region-tailored dialogues; several described experiencing perspective change. We contribute: 1) a formative study identifying goals for safe action consequence modelling, 2) the design and evaluation of a serious game integrating dialect audio, region-mapping play, bias; and 3) design implications highlighting listener-side training, transparent evaluation, and narratives maintaining psychological well-being.
\end{abstract}

\ccsdesc[500]{Human-centered computing~Empirical studies in HCI}
\ccsdesc[500]{Human-centered computing~Field studies}

\keywords{Dialect-aware Interaction, Accent Bias, Serious Games, Reflective Practice}
\begin{teaserfigure}
  \includegraphics[width=\columnwidth]{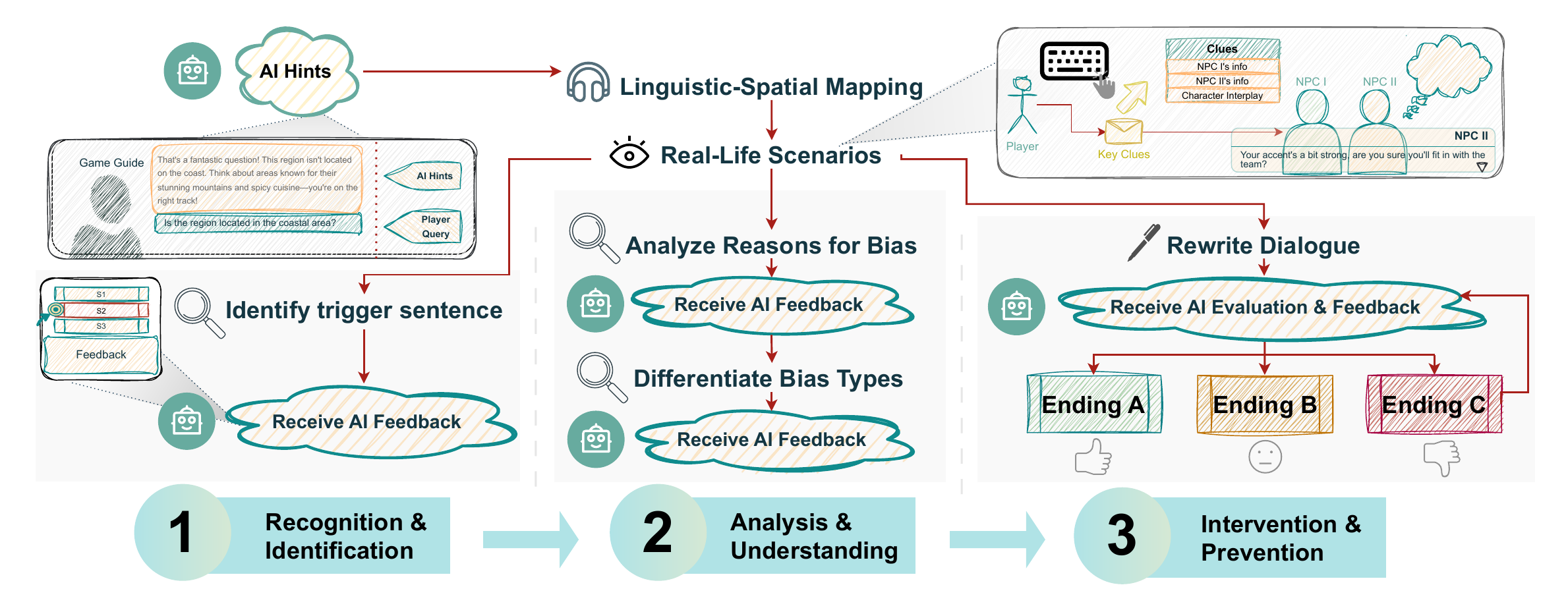}
  \caption{In this study, we designed and evaluated \textbf{CompassioMate}, a serious game aimed at mitigating dialect prejudice. The central design challenge was to create an immersive experience that tackles sensitive social issues while ensuring psychological safety and avoiding a judgmental tone. We addressed this by: embedding authentic audio and narrative context, designing three-tiered analytical progression, integrating an AI-driven dialogue modification system that provides adaptive feedback. By enabling players to actively intervene in conflicts and build their skills in a secure environment, CompassioMate fosters effective learning and critical reflection on dialect bias.}
  \Description{The image is a flow chart that visually represents the three-stage process of the serious game CompassioMate, which is designed to mitigate dialect prejudice. The flow progresses from left to right, with three numbered circles at the bottom representing each stage of the game. The first stage is represented by a user interface screen. On the left, a player profile is shown with "Dialect Detective" as the role. On the right, a dialogue box shows a conversation between the player and a non-player character (NPC), which includes an "Identify trigger sentence" button. The objective of this stage is for the player to identify the sentence that triggers bias. A speech bubble labeled "AI Hints" and a smaller dialogue box for "Feedback" are also shown, indicating that the AI provides assistance and feedback to the player in this stage. After the trigger sentence is identified, the player receives AI feedback. The second stage shows two nested processes under the heading "Real-Life Scenarios". The first process, "Analyze Reasons for Bias," involves receiving AI feedback. The second process, "Differentiate Bias Types," also involves receiving AI feedback, indicating an iterative learning loop. Above this section, there are icons and text that say "Linguistic-Spatial Mapping" and "Real-Life Scenarios," which correspond to the game's use of authentic dialect audios and customized stories. The goal of this stage is for players to understand the root causes of the bias they have identified. The final stage is where the player takes action. A dialogue box labeled "Rewrite Dialogue" is the central element. Next to it, an icon with a pen and a brush symbolizes the dialogue rewriting mechanic. After rewriting, the player "Receive[s] AI Evaluation & Feedback". This process can lead to three different narrative outcomes, labeled "Ending A," "Ending B," and "Ending C," representing different potential results of the player's intervention. This stage highlights the "AI-driven dialogue modification system that provides adaptive feedback" and enables the player to "actively intervene in conflicts and build their skills in a secure environment". The flow chart illustrates how the game moves from passive observation (Identification) to deeper cognitive engagement (Analysis) and finally to active skill-building (Intervention), all within a safe, AI-supported environment.}
\end{teaserfigure}

\maketitle

\section{Introduction} Accent bias refers to the unfair treatment or evaluation of a speaker based on how their speech sounds regardless of communicative adequacy, while dialect bias involves prejudicial judgments or decisions based on a person’s dialectal features (lexicon, grammar, phonology) that signal social identity \cite{Telo2024AccentBiasProfessionalEvaluations, Clopper2021PerceptionDialectVariation}. These related biases are not merely attitudes; they systematically shape comprehension, evaluation, and opportunity across social systems. They can lead listeners to over-attribute problems to speakers rather than contextual factors; where nonnative accents are associated with greater perceived communication difficulty and lower social belonging \cite{GluszekDovidio2010SpeakingWithANonnativeAccent, SzymanskiBrighi2025DoYouUnderstandMeCorrectly}. For children who regularly speak non-mainstream dialects, dialect–standard mismatches can hinder decoding and early reading processes, contributing to lower average literacy performance \cite{Brown2015ImpactDialect, GatlinWanzek2015DialectLiteracy, Mallinson2024LinguisticVariationInclusion}. Standard-accented candidates are rated more hirable than nonstandard-accented peers \cite{SpenceHornseyStephensonImuta2024AccentBiasHiring, Maindidze2025AccentBiasInterviews}. In addition, language barriers rooted in accent or dialect differences can increase miscommunication, leading to poorer adherence, medication errors, and lower satisfaction in healthcare \cite{AlShamsi2020LanguageBarriers, HsuehHirshMaupomeStewart2019LanguageConcordance, Diamond2019LanguageConcordance}.

Accent and dialect bias often operate beneath conscious awareness because listeners draw on rapid, automatic inferences that feel like instinctive reactions rather than prejudice. Psycholinguistic evidence shows that reduced processing fluency for accented speech reduces perceived credibility—even when the content is identical—so people sincerely experience their own skepticism as a comprehension judgment, not as bias \cite{lev-ariKeysar2010}. Social perception work similarly finds that accents trigger evaluations of status, competence, and warmth in predictable directions \cite{fuertesEtAl2012}. Crucially, such evaluations require very little speech: classic housing-audit experiments show that a few seconds of telephone speech suffice for hearers to identify dialect and make exclusionary decisions \cite{PurnellIdsardiBaugh1999}. Moreover, listeners’ expectations can manufacture accents they think they hear: matched-guise studies with teaching assistants demonstrate that merely seeing an Asian-appearing photo can lead students to report stronger accents and lower comprehension, even when the audio is standard American English \cite{Rubin1992}. Taken together, these findings explain why well-meaning people may deny holding bias: their judgments are produced by fast, automatic mechanisms that masquerade as objective assessments of clarity, intelligence, or credibility \cite{GluszekDovidio2010}.

Even when people suspect accent or dialect bias, it can be hard to acknowledge because speakers routinely preempt it by masking—co\-de-switching, style-shifting, and accommodating toward a locally standard norm. Communication Accommodation Theory holds that people adopt higher-status or in-group speech to gain approval or reduce social distance, and these normative shifts obscure the pressures driving them \cite{DragojevicGasiorekGiles2016}. Empirical research on code-switching shows that multilingual speakers adjust their phonology, lexicon, and grammar across contexts, often to manage identity and avoid negative evaluation \cite{Deuchar2020}. Crucially, preemptive masking lets both parties plausibly deny bias: the speaker can attribute convergence to audience design, while the listener can claim never to have treated anyone differently, because the stigmatized features were never present \cite{Rubin1992}. This dynamic creates a double bind. Individuals who mask succeed socially but render structural bias less visible; those who resist masking risk penalties that are rationalized as performance-based.

Accent and dialect bias are additionally difficult to detect because institutions and societies often normalize them as legitimate differences in clarity, appropriateness, or fit, reflecting a broader standard-language ideology. Scholars in linguistic anthropology describe how cultural beliefs about a singular, correct way of speaking naturalize social hierarchies and enable unequal treatment without explicitly naming it as discrimination \cite{WoolardSchieffelin1994}. Courts and workplaces have historically treated accent as a bona fide occupational qualification, allowing exclusion to be defended as quality control or customer preference, which recasts prejudice as neutral policy \cite{LippiGreen1994}. 

We introduce \textbf{CompassioMate}, an AI-mediated, dialect-aware game that transforms hidden bias into learnable, low-risk skills. This design is founded upon the principles of the Intergroup Contact Hypothesis, which posits that contact with out-group members can reduce prejudice \cite{Allport1954}. Specifically, CompassioMate utilizes indirect intergroup contact—allowing players to observe and interact with characters facing bias—to foster empathy and facilitate prejudice mitigation, a strategy particularly useful in contexts where direct intergroup contact is difficult \cite{BrownPaterson2016}. To counter the reflex of "hard to understand" with "less credible", the game includes a geolocation activity where players guess the origins of audio samples from Chinese regional varieties. This activity highlights the gaps between their accuracy and confidence while explaining how processing efforts can be mistaken for distrust. CompassioMate also addresses the masking double bind through branching narratives that let players adjust a character’s wording and stance, then observe how empathetic NPCs respond. Recognizing that this topic can be hard to discuss directly, players assume the role of a dialect detective, allowing them to uncover institutional normalization through interactive workplace and classroom scenes.

We evaluated CompassioMate through a three-week, think-aloud field study to understand how an AI-mediated serious game can help players notice, analyze, and repair everyday bias. Over time, participants advanced from simply recognizing bias to accurately labeling it and practicing wording changes that de-escalate harm. Our findings revealed that the game increased players’ ability to identify trigger utterances, distinguish bias causes, and develop concrete mitigation strategies through a dialogue-rewriting and feedback loop. Participants reported “aha” moments and stronger confidence in their compassionate communication skills. Moreover, the AI’s role as a neutral guide fostered psychological safety, encouraging candid reflection and steering players toward more inclusive alternatives. Together, these results suggest that pairing dialect geolocation play with branchable, consequence-aware conversations can transform bias from a fixed disposition into a gain in bias literacy and repair skills for everyday scenarios.

The contribution of this paper is threefold. First, we present a formative study that uncovers user expectations and pain points for bias-mitigation games, revealing the need for \textbf{learning action consequences, ensuring emotional safety, and fostering empathetic AI}. These findings inform design goals for authentic scenarios, constructive feedback, and non-prescriptive exploration, which guided the development of the full system. Second, we \textbf{develop and evaluate CompassioMate}, a serious game that integrates authentic dialect audio, region-mapping play, bias-type diagnosis, and an LLM-driven dialogue repair mechanic that guides player revisions in real time. Finally, we discussed \textbf{design implications for HCI systems that address accent and dialect prejudice}, including emphasizing listener-side training over speaker-side accommodation, providing transparent evaluation that explains why response choices succeed or fail, and branching narratives that make social consequences visible while maintaining psychological safety—principles generalizable to other microaggression and communication-repair contexts in education and work. Collectively, these contributions advance dialect-aware empathy games and provide a template for building safe, adaptive practice spaces for bias mitigation.
\section{Related Work}

In this section, we outline the motivation behind our approach. We first review existing efforts in bias mitigation, empathy, and intervention within HCI, highlighting the gaps and opportunities they present for our research. We then examine the psychological underpinnings of accent and dialect bias, aiming to understand the cognitive factors that can inform our proposed solution.

\subsection{Accent and Dialect Bias Mitigation in HCI}

Early HCI scholarship established that accent and dialect strongly influence who voice technologies serve effectively, and it introduced mitigation ideas blending sociophonetics, interface design, and breakdown–repair practices. Cambre and Kulkarni argued that the prevailing “one-voice-fits-all” paradigm marginalizes users whose accents diverge from default norms. They called for configurable, context-sensitive voices that reflect users’ linguistic diversity \cite{CambreKulkarni2019OneVoiceFitsAll}. Building on this, Cambre et al. developed a large-scale evaluation method for text-to-speech and showed that no single synthetic voice satisfies diverse listener preferences \cite{CambreMaddockTsaiColnagoKaye2020_ChoiceOfVoices}. Koenecke et al. quantified racial disparities in word-error rates across five commercial ASR systems, revealing substantially higher errors for Black speakers and underscoring the need for dialect-aware training \cite{KoeneckeNamLakeNudellQuarteyMengeshaEtAl2020_RacialDisparitiesASR}. Harrington et al. documented how Black older adults “code-switch” and negotiate identity when using voice assistants for health information, recommending culturally informed prompts, repair cues, and community-centered co-design \cite{HarringtonGargWoodwardWilliams2022_CodeSwitchingVoiceAssistant}. Mahmood and Huang further showed that voice assistants repairing accent-related errors are more effective when they take responsibility  \cite{MahmoodFungWonHuang2022_OwningMistakesSincerely}.

More recent HCI research pivots from documenting disparities to designing harm-reduction and dialect-aware tools, increasingly using LLMs and socio-technical evaluation. Wenzel et al. showed that ASR errors act as microaggressions for Black users and proposed asset-based designs that include affirming acknowledgments during error repair \cite{WenzelDevireddyDavidsonKaufman2023_CanVoiceAssistantsBeMicroaggressors}. Michel et al. found accent bias embedded in cloned voices, arguing for provenance transparency and identity-reflective voice choices at design time \cite{michel2025accent}. Ezema et al. showed how ASR bias cascades into downstream inequities in spoken-interface tutoring \cite{ezema2025biasASR}. Finally, Egede et al. compared a culturally informed LLM assistant (ChatBlackGPT) with a general-purpose LLM, showing that community-centered, dialect-aware LLMs can better align with Black users’ values and tasks while raising new questions about evaluation and prompting for inclusive design \cite{egede2025chatblackgpt}.

Prior studies largely focus on documenting bias, auditing systems, or prescribing reactive fixes such as code-switching or apology protocols after errors. This leaves limited space for proactive, experiential approaches that offer safe, low-stakes environments where people can explore inclusive design strategies, learn from mistakes without harm, and iteratively build dialect-aware skills. Existing methods rarely allow users to experiment with alternative wordings or observe how subtle social cues shape real-time interpersonal dynamics. While valuable, these approaches leave limited space for proactive, experiential learning.  Addressing these gaps calls for a novel tool that combines immersion, adaptive feedback, and reflective practice to transform bias awareness from passive observation into active, situated learning.

\subsection{Empathy and Intervention in HCI}

Recent HCI research increasingly explores LLM–mediated empathy and interventions, aiming to support perspective-taking and prosocial outcomes while addressing ethical constraints and promoting safe, non-shaming reflection. Cuadra et al. (2024) examined how LLM-powered conversational agents can appear empathetic while lacking true understanding \cite{cuadra2024illusion}. Wu et al. (2024) introduced MindShift, a just-in-time intervention that uses LLMs to generate context-aware, goal-aligned persuasive messages based on users’ mental states \cite{wu2024mindshift}. Jörke et al. (2025) presented GPTCoach, an LLM-based health-coaching probe that applies motivational interviewing techniques—posing open-ended questions, normalizing ambivalence, and co-developing plans from sensor data \cite{joerke2025gptcoach}. Ju et al. (2025) proposed EmoSync, which crafts personalized analogical scenarios to help users empathize with others’ microaggression experiences through familiar contexts \cite{ju2025toward}. Dongre et al. (2024) outlined physiology-driven EmLLMs that integrate wearable sensing with LLM dialogue to adapt empathic responses in real time \cite{dongre2024physiology}. Zhang et al. (2024) introduced ProcrastiMate, a text-based serious game designed to promote non-shaming self-reflection on procrastination. In the game, players take on the role of a counselor who identifies the causes of procrastination and explores coping strategies \cite{zhang2025walk}.
 Empirically, gamified interventions have proven highly effective in mitigating  cognitive biases: gamification of cognitive bias  significantly increases user engagement necessary to change interpretation biases in anxiety \cite{Salemink2022}, and it successfully helps users avoid complex cognitive biases, such as judgmental forecasting \cite{Legaki2021}. 

Despite these advances, important gaps remain that point to opportunities for future work. Existing approaches often emphasize individual reflection, emotional regulation, or behavior change in relatively isolated settings, but rarely address how subtle, everyday social biases shape real-time interpersonal dynamics. Specifically, they tend to overlook low-stakes, socially situated practice spaces where people can experiment with different responses and observe their emotional impact on others without fear of judgment. This points to the unique potential of LLM-driven serious games that combine playful interaction with reflective learning, allowing users to explore branching conversational scenarios and witness how small shifts in tone or wording influence social outcomes.

\subsection{Understanding Accent and Dialect Bias}

Accent and dialect bias are often driven by external factors and raciolinguistic ideologies that dictate which ways of speaking are valued or stigmatized. This systemic process is clarified by the \textit{language subordination framework}, which posits that societal institutions reinforce unequal power relations by establishing and upholding language hierarchies \cite{LippiGreen2012}. Specifically within the Chinese context, these ideologies have deep historical roots in the national language project, which systematically established Mandarin hegemony and codified local \textit{fangyan} (dialects) into a fixed hierarchy of prestige \cite{Tam2020}. In schools and workplaces, standardized language is upheld as the “proper” norm, listeners often racially frame speakers as deficient even when they use that norm, and institutional policies like academic language standards reinforce these inequities by rewarding code-switching and penalizing dialect features \cite{flores2015undoing}. 

There have been institutional changes proposed such as shifting intervention targets from speakers to listeners, standardizing evaluation practices, and adopting linguistically inclusive pedagogy and policy. First, research emphasizes listener-focused training that builds sustained exposure to diverse accents, which reduces processing effort and improves social judgments of speakers, suggesting that accent bias often stems from unfamiliarity rather than inherent deficit \cite{RovettiSumantryRusso2023, BradlowBent2008, BoduchGrabkaLevAri2021}. Complementary accessibility measures, such as same-language captions and transcripts, further support comprehension and recall \cite{Gernsbacher2015}. Second, organizations and schools can minimize bias in evaluations by adopting structured, job-relevant assessments. Evidence from personnel psychology demonstrates that structured interviews—with standardized prompts, anchored rubrics, and trained raters—are more valid and resistant to bias than unstructured formats \cite{LevashinaHartwellMorgesonCampion2014}. These practices also increase fairness and diversity outcomes in selection processes \cite{BergelsonTracyTakacs2022}. Third, linguistically inclusive pedagogy, such as code-meshing and translanguaging, affirms students’ full repertoires while teaching disciplinary conventions. Policy statements like the Conference on College Composition and Communication’s \textit{Students’ Right to Their Own Language}, alongside systematic reviews, show that such approaches yield positive academic and affective outcomes when evaluation emphasizes ideas over surface dialect features \cite{CCCC_StudentsRightOwnLanguage_2014, Prilutskaya2021}.

A body of research has examined the psychological and personal causes of dialect bias. Four interrelated causes have been identified. 1) \textit{Social prejudice}: Accents can mark speakers as members of an out-group, triggering stereotypes about “outsiders.” This leads to reduced credibility judgments based on social bias rather than message content. Listeners often map accents onto perceived status cues such as intelligence or social class, which results in more positive evaluations of “standard” varieties and penalties for minority or non-standard accents \cite{lev-ariKeysar2010}. 2) \textit{Solidarity signaling}: Beyond prejudice, accents function as identity markers that influence judgments of in-group versus out-group membership. These perceptions affect evaluations of warmth, trustworthiness, and likeability even when identical information is conveyed, showing that bias often stems from social identity processes \cite{fuertesEtAl2012}. 3) \textit{Low processing fluency}: Accented speech can require more cognitive effort to understand. This extra effort and processing difficulty lead listeners to misattribute the difficulty to the speaker’s credibility or the truthfulness of the content. As a result, identical statements are often judged less accurate or persuasive when spoken with an accent, even in the absence of explicit prejudice \cite{lev-ariKeysar2010, fuertesEtAl2012}. 4) \textit{Unfamiliarity}: When listeners lack exposure to a dialect, comprehension challenges emerge. This unfamiliarity can unfairly translate into negative judgments of competence or professionalism \cite{rickford2016language}.

Recognizing these mechanisms in concrete cases can drive improvement, as awareness combined with targeted strategies reframes bias from a fixed trait into a correctable habit. For example, interventions that teach individuals to notice cues of stereotyping, generate concern, and practice evidence-based techniques (e.g., perspective taking, counter-stereotypic imaging) have been shown to produce lasting reductions in implicit bias months later \cite{DevineForscherAustinCox2012}. Moreover, when listeners understand that reduced credibility judgments may stem from processing difficulty rather than speaker competence, they can correct for this misattribution \cite{lev-ariKeysar2010}. Exposure itself also helps: even a minute of experience with accented speech accelerates comprehension, indicating that training paradigms can rapidly reduce bias-driven responses \cite{ClarkeGarrett2004}. Serious games are effective vehicles for such metacognitive and practice-based learning, yielding small-to-moderate gains in knowledge and retention across domains \cite{WoutersVanNimwegenVanOostendorpVanDerSpek2013} and increasing bias literacy through perspective taking \cite{KaatzCarnesGutierrezSavoySamuelFilutPribbenow2017}. This is supported by game-based interventions, where positive parasocial contact with outgroup characters has been shown to increase empathy and reduce stereotypes \cite{chen2022effect}.

This approach is fundamentally grounded in the \textbf{Intergroup Contact Theory} proposed by Allport (1954) \cite{Allport1954}, which posits that contact between members of different groups reduces prejudice. Extensive meta-analytic research confirms that intergroup contact typically reduces prejudice across diverse settings and outgroup targets, with the effect often generalizing to the entire outgroup \cite{PettigrewTropp2006Contact}. Furthermore, Wright et al. (1997) have established that this effect extends beyond direct, face-to-face interactions to include \textbf{indirect (or extended) contact} \cite{Wright1997}. Research demonstrates that indirect contact (e.g., knowing an ingroup friend with an outgroup friend) reduces prejudice at levels comparable to those of direct contact, with both forms enhancing the overall prediction of prejudice reduction through the mechanism of establishing tolerant intergroup norms and reducing intergroup threat and anxiety, although direct contact is uniquely effective at reducing both collective and individual threat, whereas indirect contact is only slightly related to reducing individual threat \cite{PettigrewChristWagnerStellmacher2007}.

\section{Research Aim}
In summary, our research aims to \textbf{mitigate accent-based prejudice} from an educational perspective, utilizing serious games as a medium. This overarching goal is comprised of three specific objectives: 1) to identify the key challenges and objectives in creating a serious game that reduces accent and dialect bias; 2) to design and implement a game based on these objectives; and 3) to test the learning process, outcomes, and player experience, thereby contributing to the HCI community's understanding of this domain.

To achieve these objectives, we first clarify the theoretical foundation to be integrated into the game. We grounded our game design in the 4 root causes of accent bias, which, as discussed in subsection 2.3: 1) Social prejudice, 2) Solidarity signaling, 3) Low processing fluency, and 4) Unfamiliarity. Additionally, we identified 3 major manifestations of accent bias through an extensive literature review. Both the root causes and manifestations guided our subsequent game design decisions. We then adopted a \textbf{Research through Design} approach. We first conducted a formative study (section 4) using the initial game elements to determine the design goals for the full version of the game. Following the design and implementation of our serious game (section 5), CompassioMate. Finally, we detailed our evaluation approaches (section 6) which employed a Think-Aloud Protocol, and present our findings (section 7).
\section{Formative Study} \label{sec:formative}
To address our first research objective, we organized a formative study targeting university students as potential players. We selected this group not only due to their high familiarity with regional dialect culture and internet slang \cite{zheng2024analysis}, but strategically because they represent a \textbf{pivotal social group} whose current language practices will shape the future of Chinese society \cite{Jiang2018}. Furthermore, prior research indicates that young, highly-educated adults report \textbf{significantly higher anxiety and lower confidence} in the use of Chinese dialects \cite{Jiang2018}, making them a critical target audience for an intervention aimed at fostering dialect awareness and empathy. In this section, we will present our initial design elements, research methods, and key findings from student participants and experts, leading to our three new design goals.

\subsection{Initial Design}
Inspired by the game \textit{GeoGuessr}\footnote{\url{https://www.geoguessr.com/}}, our game design also emphasizes problem-solving through geolocation. Players will identify the speaker's origin by listening to audio recordings of their accent and then select the corresponding region on a map of China. Specifically, our initial game design aimed to transform the abstract task of accent recognition into a concrete geolocation challenge. We employed two key design elements:

\begin{enumerate}
    \item \textbf{Authentic Dialect Audios} played while players select a region on the map. A total of 150 audio recordings were sourced from the China Language Resources collecting and recording platform \footnote{\url{https://zhongguoyuyan.cn/}}. Representative dialect points were selected from each sub-region within the seven major dialect areas: Wu, Yue, Min, Hakka, Gan, Xiang, and Mandarin. For example, in the Yue dialect area, Guangzhou was selected to represent the Guangfu sub-region, while Dongguan represented the Guanbao sub-region. A total of 21 speakers were selected from each dialect area except for Mandarin, totaling 126 ($21 \times 6$) recordings. Due to the vast geographical coverage of the Mandarin (Guanhua) area, one representative point was selected from each of its eight sub-regions—such as Southwestern Mandarin, Northeastern Mandarin, and Jianghuai Mandarin—with recordings from three speakers per point, totaling 24 ($8 \times 3$) recordings.
    \item \textbf{Dialect Prejudice Scenarios} presented through short stories. We use brief stories to guide players into diverse accent bias scenarios. These stories are told in the third person, with plots shaped by the social, social-psychological, and cognitive roots of accent prejudice. The settings are ordinary and everyday—offices and schools—for example, in a job interview scenario, being deemed unqualified due to an accent from a less developed region, to enhance personal connection.
\end{enumerate}

Based on the above game elements, a preliminary gameplay sequence was developed to ensure the game experience. The game proceeds as follows: 1) Players listen to audio recordings containing accents. 2) They select a region on the map based on the accent's characteristics and are informed of the natural and cultural fun facts of the selected province, as well as the geographical distance between their selected province and the correct province. After selecting the correct region, players proceed to an everyday story scenario. 3) After watching the storyline, players need to reflect on the scenario and select the cause of the NPC's accent bias. Incorrect selection results in the game providing an explanation for the wrong option, and the player continues until they select correctly. 4) The game randomly draws from 24 audio segments for playback, and the game continues until players complete all 8 game rounds.

\subsection{Methods}
Employing the initial gameplay, we conducted a guided Cognitive Walkthrough (CW) of the game prototype with 16 potential players to gather feedback on our design. This section details participant recruitment, setup, procedures, data collection, and data analysis. The study received ethical approval from the first author’s institution.

\subsubsection{Participants}
We recruited participants from a public forum at a university in China, seeking individuals who considered themselves interested in dialects and willing to explore the topic further. Our study included 16 participants (G1 - G16; 7 males, 9 females). All participants were university students aged 18–24. Each participant who completed the study received a ¥50 coffee voucher.

\subsubsection{Setup}
The study was conducted in a private, quiet room at the first author’s institution to minimize distractions and create a comfortable environment for participants. The room was equipped with a desktop computer with a high-resolution monitor and a microphone to capture participants' verbal comments. The game prototype, developed using the Godot engine \cite{godot_engine}, was pre-loaded on the moderator's computer and ready for the guided walkthrough.

The setup was designed to facilitate a CW while capturing user feedback. Although CW is typically an expert-based method performed by evaluators to assess a system's ease of learning through exploration \cite{lewis1990testing, polson1992cognitive}, our study adapted this approach by having the first author perform the walkthrough with a group of potential users.

To ensure comprehensive data collection, the following was implemented:
\begin{itemize}
    \item \textbf{Screen and Audio Recording:} Screen Studio was used to record both the computer screen and the audio from the room. This captured the visual flow of the game prototype as well as the participants' uninhibited verbal comments.
    \item \textbf{Facilitator's Role:} The first author, acting as the moderator, guided the walkthrough. The moderator verbally explained the game's mechanics and design rationale, while the participants were instructed to vocalize their thoughts, questions, and feedback as they watched the demonstration. This hybrid approach allowed for the application of CW principles in a user-centric context.
    \item \textbf{Data Synchronization:} A synchronized timestamp was used across all video and audio recordings to ensure that verbal feedback could be precisely matched to the specific gameplay moment it referred to during data analysis.
\end{itemize}
All collected data, including the video recordings and field notes, was securely stored on an encrypted local drive. This ensured the confidentiality and integrity of the data, accessible only to the research team.

\subsubsection{Procedures and Data Collection}
We began each session by explaining the study’s purpose and procedures, emphasizing that our goal was to evaluate the game prototype, not their personal abilities. Participants were informed about the data collection process, including the use of screen and audio recordings, and were assured of the confidentiality of their data. Each participant signed a consent form before the session began.

Our procedures were guided by the principles of Cognitive Walkthrough (CW), a method that evaluates a system's ease of learning through exploration\cite{jadhav2013usability}. While CW is traditionally performed by an expert evaluator assessing a predetermined sequence of actions, we adapted this method to a user-centric approach. We applied the four key questions of a CW \cite{wharton1994cognitive} to our hybrid session, observing the participants' experience directly. We assessed whether participants would: 1) try to achieve the right effect, by observing if they understood the game's objectives (e.g., identifying the correct dialect region on the map); 2) know that the correct action is available, by noting if they could identify interactive elements; 3) associate the correct action with the intended effect, by assessing if they connected on-screen actions (e.g., clicking on a province) with the intended outcome (e.g., confirming their guess); and 4) recognize if progress is being made, by observing whether they recognized the system's feedback.

The session unfolded in the following steps. Firstly, after obtaining consent, the first author provided a brief overview of the game's concept, its anti-prejudice goals, and the gameplay sequence. Secondly, acting as the moderator, the first author guided participants through a fixed sequence of game scenarios, including playing the dialect audio, demonstrating map interaction, and presenting short prejudice scenarios. Thirdly, as the walkthrough progressed, participants were encouraged to think aloud and vocalize their thoughts, questions, and feelings. The moderator captured these real-time comments. Lastly, participants were asked to complete our Pilot Study Survey containing  open-ended questions.

All sessions were recorded with Screen Studio, capturing both screen activity and audio. The synchronized timestamps allowed us to link specific verbal feedback to the corresponding moments in the walkthrough. The moderator also took written notes to capture additional observations.
\subsubsection{Data Analysis}
The analysis of the open-ended data (including video transcripts and survey responses) proceeded via inductive approach to address Research Objective 1. Two independent researchers performed open coding on the raw textual data, generating granular codes based on participants' experiences, expectations, and observations of accent bias. Subsequently, the researchers utilized affinity diagramming \cite{lucero2015using} to synthesize these codes into three overarching themes. For instance, observations such as 'judgment in job interviews' and 'being corrected by peers' informed the theme \textit{Game Content and Learning Experience}. Emotional descriptors like 'shame' and 'stress avoidance' underpinned \textit{Player Identity and Emotional Resonance}, while insights regarding 'empathy' and 'new knowledge' led to the theme \textit{Game Appeal and Expectations}. These themes are elaborated in the following findings section.
\subsection{Findings}
\subsubsection{Game Mechanics Guide Players to Confront and Reflect on Bias}
The data presented below reveals how the game's narrative, gameplay, and interactive mechanics directed players' attention toward and prompted reflection on real-world prejudices against dialects (e.g., "dialects are backward," "dialects hinder communication") and linguistic diversity. Participants expressed that the game's task setup enabled them to experience deep empathy and a strong sense of mission. Many agreed with the game's intent to protect dialect culture and lamented the gradual disappearance of regional accents. For example, G13 described feeling "empathetic because I have indeed experienced the gradual disappearance of dialects among the younger generation in my life… I believe they shouldn't be abandoned as intangible cultural heritage." The game also increased their attention to the issue of accent bias, with G2 stating, "Now I realized that many people are prejudiced against accents."

\subsubsection{Authentic Scenarios Foster Empathy and Awareness}
The perceived fidelity of the game's scenes, dialogues, and choices to participants' real-life experiences correlated with high self-reported engagement and emotional resonance. The majority of participants confirmed this connection, noting that the scenarios closely resembled real-life situations. For example, G1 described a scenario where prejudice arose due to accent differences: "It makes me recall experiences of communicating with classmates from different regions in real life." The game also prompted significant attitude shifts toward accents, with some participants indicating that the game made them start caring about the issue of accent bias. By empathizing with the NPCs in the game, they gained a clearer understanding of the regional prejudice that some accent speakers face in society today, as G4 described, "What's even more frightening is that one's hometown can be identified by their dialect, leading to prejudice based on regional stereotypes.

\subsubsection{Player Experience and Suggestions for Game Improvement}
Many players feel this game is an innovative and effective way to learn dialects and cultural knowledge through engaging interactions. G5 found the \textit{GeoGuessr}-style gameplay, where players identify locations by accent and get fun facts, particularly interesting. Participants showed enthusiasm not just for dialect diversity ("New linguistic knowledge! I really want to learn about more accents!" - G15), but also for understanding the reasons behind dialect bias. As G6 noted, "What's great is that this game uses dialect as a starting point to help players understand the underlying cultural and social reasons for prejudice, allowing me to learn a lot."

However, some players felt that while the game has educational value, its "edutainment" approach still has room for improvement. Some players pointed out that the game felt more like educational software, with slightly lacking gameplay, potentially making it less appealing than purely fun games. They believed that if it were just a "reflection segment," it might bore some players, especially when the answers were preset by the system, lacked substantial impact, or personalized output. This could be perceived as passive didacticism, leading to a lack of "satisfaction." Therefore, it was suggested that the game could be more personalized or "lively," increasing interactivity with player tags to enhance the individual experience.

\subsection{Summary of Design Goals}
Based on the findings in Section 4.3, we summarized that the prototype achieved strong emotional resonance and promoted active reflection, but this success came at the cost of the user’s sense of agency. Specifically, the qualitative data highlighted a trade-off: Didactic clarity reduced player choice satisfaction. Consequently, we formalized new design goals for our full game, aiming to sustain the emotional impact while significantly enhancing player agency and personalization, directly addressing the limitations identified in the pilot study.

\aptLtoX{\begin{itemize}
    \item[\textbf{[{DG1}]}]\label{dg:1} \textbf{To Deepen Emotional Connection and Personal Reflection:} By immersing players in authentic, emotionally resonant, and subtle scenarios of accent-based bias, the game will foster a deep emotional connection, enhance personal relevance, and actively prompt critical reflection on prejudice that might otherwise go unnoticed. This approach ensures players personally understand the human cost of bias through vicarious experience and critical self-examination.
    \item[\textbf{[{DG2}]}]\label{dg:2} \textbf{To Guide Objective Exploration for Optimal Communication Outcomes:} The game will encourage players to engage with bias scenarios from an analytical third-person perspective. This detachment allows players to move beyond negative emotions (like stress or discomfort) to objectively diagnose the deeper reasons behind prejudice. Through constructive and structured feedback, the game will guide players toward effective, optimal solutions for real-life communication challenges.
    \item[\textbf{[{DG3}]}]\label{dg:3} \textbf{To Facilitate Fluid Knowledge Integration and Player Agency:} By utilizing a dynamic dialogue system (generative dialogue), the game will provide feedback that is fluid, natural, and non-prescriptive. This system promotes a strong sense of agency by responding directly to the player's choices, thereby making the learning process feel natural, personally meaningful, and ensuring a smooth, effective integration of communication knowledge.
\end{itemize}}{\begin{designgoals}
    \item\label{dg:1} \textbf{To Deepen Emotional Connection and Personal Reflection:} By immersing players in authentic, emotionally resonant, and subtle scenarios of accent-based bias, the game will foster a deep emotional connection, enhance personal relevance, and actively prompt critical reflection on prejudice that might otherwise go unnoticed. This approach ensures players personally understand the human cost of bias through vicarious experience and critical self-examination.
    \item\label{dg:2} \textbf{To Guide Objective Exploration for Optimal Communication Outcomes:} The game will encourage players to engage with bias scenarios from an analytical third-person perspective. This detachment allows players to move beyond negative emotions (like stress or discomfort) to objectively diagnose the deeper reasons behind prejudice. Through constructive and structured feedback, the game will guide players toward effective, optimal solutions for real-life communication challenges.
    \item\label{dg:3} \textbf{To Facilitate Fluid Knowledge Integration and Player Agency:} By utilizing a dynamic dialogue system (generative dialogue), the game will provide feedback that is fluid, natural, and non-prescriptive. This system promotes a strong sense of agency by responding directly to the player's choices, thereby making the learning process feel natural, personally meaningful, and ensuring a smooth, effective integration of communication knowledge.
\end{designgoals}}

\section{Design and Implementation of CompassioMate}
To address our second research objective, we developed CompassioMate, a text-based adventure game whose design was directly informed by our formative study. The game's implementation was guided by the resulting design goals, leading to three design considerations: embedding authentic audio and narrative context, designing three-tiered analytical progression, integrating an AI-driven dialogue modification system that provides adaptive feedback.

\subsection{Game Settings}
\subsubsection{Player’s Role and Storyline}
\begin{figure*}
\includegraphics[width=\linewidth]{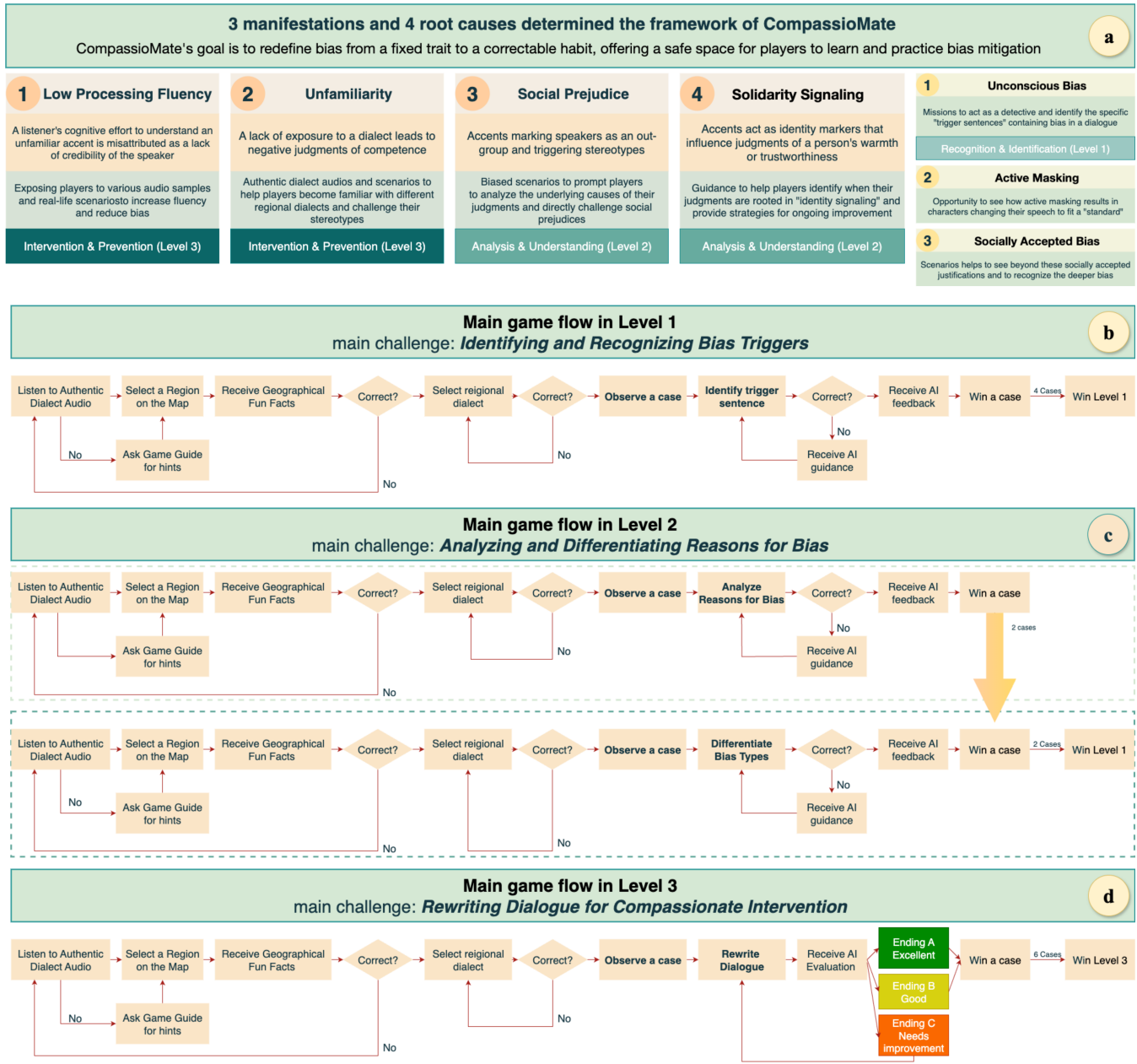}
\caption{Overview of CompassioMate. a) The game's core design is grounded in the four root causes and three manifestations of accent bias that we identified. b) Level 1's core gameplay mechanics challenge players to identify and recognize bias triggers. c) Gameplay in Level 2 requires players to analyze and differentiate reasons for bias. d) Gameplay in Level 3 challenges players to rewrite dialogues to intervene and resolve conflicts.}
\label{fig:flow}
\Description{his figure is a comprehensive, multi-part diagram that visually outlines the design framework and progressive gameplay of the CompassioMate serious game. It consists of four distinct sections, labeled (a), (b), (c), and (d), which detail the game's theoretical underpinnings and its three-level progression. Section (a): Framework and Key Concepts This top section establishes the game's core design, which is grounded in the four root causes and three manifestations of accent bias that we identified. It is a grid of seven labeled boxes. The four root causes are presented horizontally as: 1) Low Processing Fluency: Defined as "a lack of acoustic credibility of the speaker. 2) Unfamiliarity: Described as "a lack of exposure to a dialect leads to negative judgments of competence.” 3) Social Prejudice: Explained as "biased scenarios to prompt players to reflect on their own judgments and directly challenge social prejudices." 4) Solidarity Signaling: Defined as "influence judgments of a person's warmth or trustworthiness." The three manifestations are listed vertically on the far right as: 1) Unconscious Bias: Missions to act as a detective to identify the specific trigger. 2) Active Masking: Opportunities to use their skills to blend in. 3) Socially Accepted Bias: Players learn to recognize justifications and "the deeper bias." The diagram connects these concepts to the specific levels of the game, showing a clear theoretical basis for the gameplay. Section (b): Main Game Flow in Level 1 This section, a detailed flowchart, describes the gameplay progression for Level 1, where the main challenge is "Identifying and Recognizing Bias Triggers." The flow begins with the player's action to "Listen to Authentic Dialect Audio" and then "Select a Region on the Map." A side path allows the player to "Ask Game Guide for hints." If the player's selection is "Correct?", they receive "Geographical Fun Facts." The flow then leads to "Observe a case," which is the central event. Here, the player is challenged to "Identify trigger sentence." If their identification is "Correct?", they receive "AI feedback," win a case, and progress. If not, they receive "AI guidance" and loops the player back to the task, reinforcing the learning. Section (c): Main Game Flow in Level 2 This flowchart details the gameplay for Level 2, with the main challenge being "Analyzing and Differentiating Reasons for Bias." The flow is similar to Level 1, starting with listening to audio and selecting a region. After "Observe a case," the core actions require the player to first "Analyze Reasons for Bias" and then "Differentiate Bias Types." For both of these steps, a "Correct?" decision point leads to "Receive AI feedback," winning a case, and progression. An "Incorrect" response provides "AI guidance" and loops the player back to the task, reinforcing the learning. Section (d): Main Game Flow in Level 3 The final flowchart illustrates Level 3, with the main challenge being "Rewriting Dialogue for Compassionate Intervention." The flow begins with the familiar steps of listening to audio, selecting a region, and observing a case. The central action is to "Rewrite Dialogue," which challenges players to apply all the skills they have learned. After this action, they "Receive AI Evaluation." This final step leads to one of three possible outcomes: "Ending A (Excellent)," "Ending B (Good)," or "Ending C (Needs Improvement)." A successful outcome (A or B) contributes to winning a case and successfully completing the game's final objective.}
\end{figure*}

CompassioMate is set in a virtual society where players assume the role of a secret accent detective. Their primary task is to identify the geographical region corresponding to various accents and discern the trigger phrases and underlying causes of accent bias in different scenarios. Furthermore, the game empowers players to alter the story's outcome by intervening in developing "cases"—disputes or unpleasant situations sparked by accent discrimination. To prevent the escalation of conflict, players must identify and intervene in suspicious dialogue from the source of the prejudice. To successfully complete the game, players must perform well on every task, resolving all 14 (4+4+6) game cases. This multi-level structure, as illustrated in Figure \ref{fig:flow}, is grounded in the three manifestations and four root causes of accent bias that our literary review revealed, and is informed by the \textit{psychosocial moratorium principle}, which suggests that digital learning spaces enable individuals to safely practice difficult skills because the consequences of failure are minimized \cite{Gee2003}.

\subsubsection{Linguistic-Spatial Mapping}
This foundational game setting immerses players in a low-stakes, interactive phase designed to link auditory perception with geographical knowledge. In this stage, players listen to authentic dialect audio and identify its corresponding region on a map. An interactive AI guide provides non-critical clues, encouraging exploration. This process, grounded in a non-judgmental feedback loop, encourages players to build cognitive associations between sound and place, preparing them for the more complex challenges ahead.

\subsubsection{Customized Accent Bias Stories}
We prioritized creating custom stories tailored to the user background, combining academic theory with familiar life experiences.

\textbf{Situational Specificity Enhances Contact}: Both the game's scenarios and the authentic, nuanced dialogue content were meticulously curated based on the inductive analysis of our pilot study data (as described in section 4.2.4), which systematically identified granular codes for participants' real-life experiences and observations of accent bias scenarios. This reliance on familiar, real-world context directly supports the goal of \textbf{indirect contact} by increasing the realism and relevance of the vicarious experience \cite{BrownPaterson2016}.

\textbf{Theory-Driven Scenario Structure}: Building upon the \textit{Intergroup Contact Theory} \cite{Allport1954}, the design aims to facilitate this indirect contact by establishing scenario conditions conducive to reducing intergroup bias, with each narrative crucially constructed to instantiate the theory's classic facilitating conditions (equal status, common goals, and institutional support) to maximize the psychological effect of this simulated contact. For instance, reflecting the common stressor of team assignments, we developed a scenario where students collaborating on a complex Maths project in a library experience communication friction when an unfamiliar accent leads to one student’s unintentional exclusion from a key decision.

\begin{figure*}
    \centering
    \includegraphics[width=\linewidth]{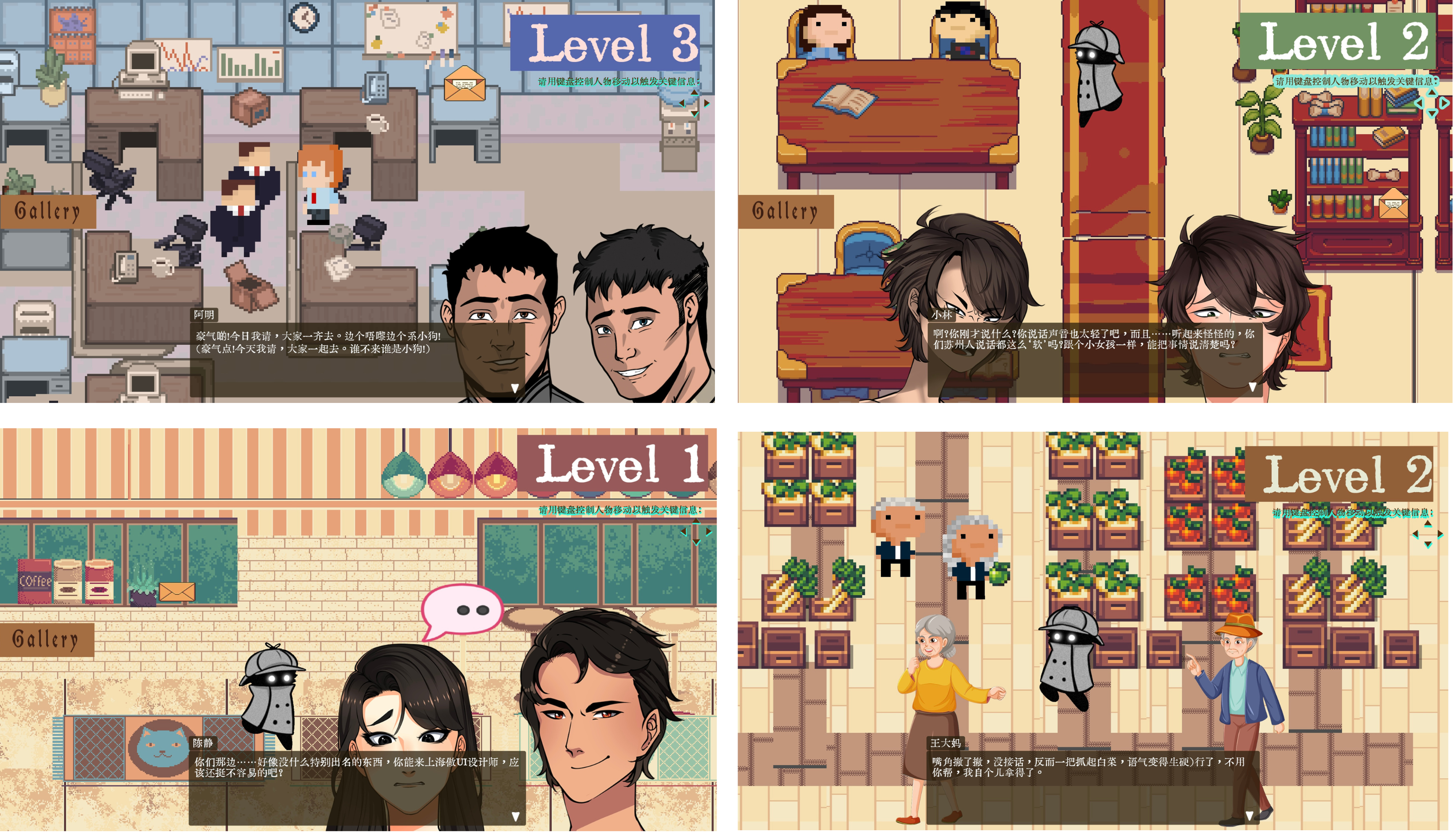}
    \caption{Visual examples of the game’s interactive world and its character-driven narrative}
    \label{fig:scenes}
    \Description{This single image, divided into four panels, shows the interactive world of a serious game with a character-driven narrative. Rendered in a 2D cartoon style, each panel showcases a different realistic social scenario. Players control detective characters who navigate and interact with the world through a dialogue interface. The top-left panel is set in a coffee shop, where a character with a backpack stands opposite a barista, with a dialogue box prompting the player to "Identify trigger sentence." The top-right panel shows a modern office setting with two characters, likely colleagues or an interviewer and candidate, engaged in a conversation. The bottom-left panel depicts two students seated at a desk in a university library, representing a school-based scenario. Finally, the bottom-right panel shows a scene in a grocery store with three characters and a dialogue box, likely representing a tense social situation. In each panel, the dialogue box highlights that players will be engaging in conversations to progress the narrative.}
\end{figure*}

In the game world, players can control the detective’s movement using the keyboard, triggering conversations as key NPCs approach. Figure \ref{fig:scenes} illustrates some typical game scenes, showcasing the interactive environment and character interactions.

\subsection{Level 1: Recognition \& Identification}
\begin{figure*}
    \centering
    \includegraphics[width=\linewidth]{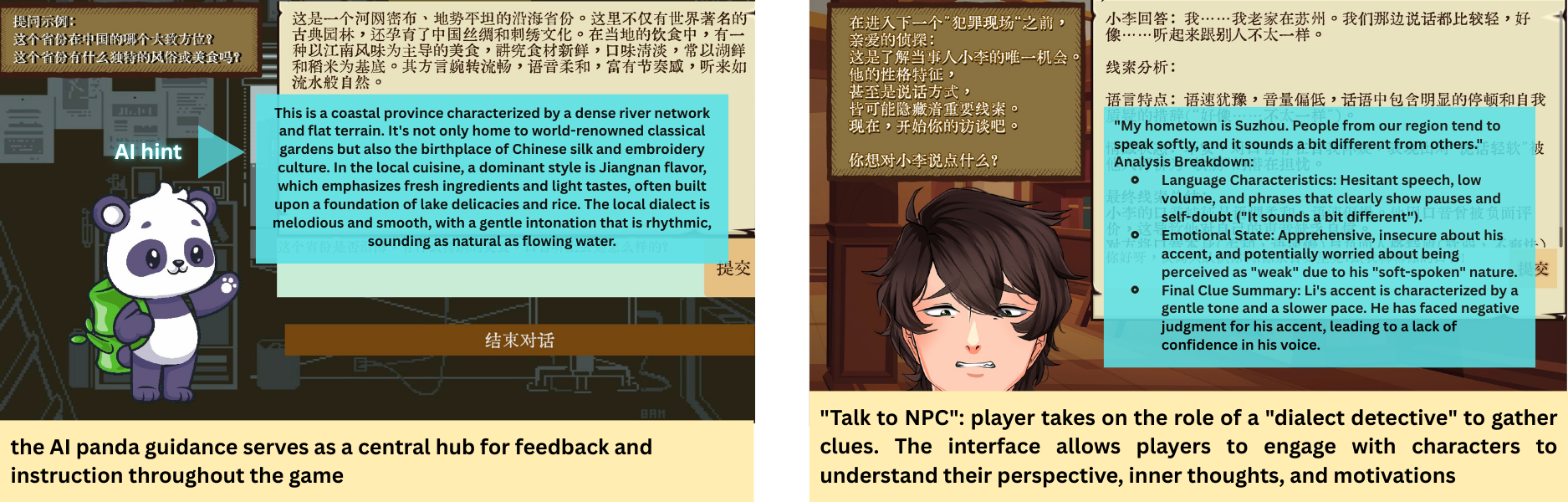}
    \caption{The game's AI system is comprised of two core components: the AI Panda Guide, which provides objective feedback, and the AI-driven NPCs, which enable players to gather clues and understand characters' inner thoughts.}
    \Description{The game's AI system is central to its learning objectives, operating through two main components. The AI Panda Guide serves as the primary feedback mechanism, designed as a "neutral referee" to ensure a psychologically safe learning environment. It provides real-time, adaptive feedback that helps players navigate sensitive social scenarios without feeling judged. This functionality is crucial for guiding players through the game's three levels, from identifying bias triggers to analyzing their root causes and, finally, to intervening with a compassionate response.The second component consists of the AI-driven NPCs, which are integral to the game's character-driven narrative. The "Talk to NPC" page allows players to act as a "dialect detective" to engage with characters, uncover clues, and learn about their personal histories and emotional responses to bias. This interaction is designed to make the scenarios feel realistic and to foster empathy by encouraging players to understand the situation from the characters' perspectives. By interacting with the AI NPCs, players gain the context and motivation needed to effectively use the skills they learn from the AI Panda Guide.}
    \label{fig:ai}
\end{figure*}

Level 1 serves as a foundational buffer stage before Levels 2 and 3 begin. Through dialogue and interaction with the game's guides—a panda spirit and core NPCs—players are informed of their role, receive encouragement, accept the game's mission, and begin the first round of gameplay. In Figure \ref{fig:ai}, we see the AI guides' role in providing feedback and clues to the player.

\paragraph{Gameplay in Each Case}
First, players enter the most basic stage, "Geoguesser," where they listen to diverse dialects and select the corresponding geographical region on a map to form a general impression and association. In the four cases of Level 1, after successfully locating the accent's region, players proceed to identify the bias trigger. They enter an everyday scene, control the detective's movement via the keyboard, and trigger a dialogue by approaching a key NPC. After observing the conversation, players select the trigger sentence containing accent discrimination from all "suspect" dialogues. Regardless of whether their choice is correct, the game guide provides gentle and personalized feedback from bias intervention literature.

\subsection{Level 2: Analysis \& Understanding}
After the introductory warm-up in Level 1, players are informed that after correctly identifying the "case clues" (biased language), they must find the "motive" behind these clues.

\paragraph{Gameplay in Each Case}
After identifying the dialect's geography and observing a dialogue, players must pinpoint the underlying reason for an NPC's accent bias. The first stage requires correctly identifying four root causes of bias. The game's complexity increases in Level 2, where players must then recognize three more subtle and challenging manifestations of accent bias to complete the level.

Throughout this process, the panda guide provides immediate and personalized feedback. An incorrect selection triggers a detailed explanation of why that bias type is mismatched, whereas a correct choice elicits a positive response. This feedback mechanism is instrumental in cultivating players' critical thinking skills, encouraging a rational rather than emotional approach to the intricate issue of accent bias. Specifically, we aim to prevent defensive emotional tactics (such as guilt, denial, or avoidance) that are shown to impede learning and resistance to disruptive experiences \cite{Gillespie2020}.

\subsection{Level 3: Intervention \& Prevention}
After completing the challenges of Levels 1 and 2, players are expected to have developed the ability to correctly identify and understand accent bias. Building on this foundation, the game escalates the task to a more complex stage: \textbf{dialogue repair and reconstruction}. This direct intervention approach aligns with literature demonstrating that intentional strategies can effectively reduce implicit bias \cite{Lai2014ReducingBias}.

\begin{figure*}
  \centering
  \includegraphics[width=\linewidth]{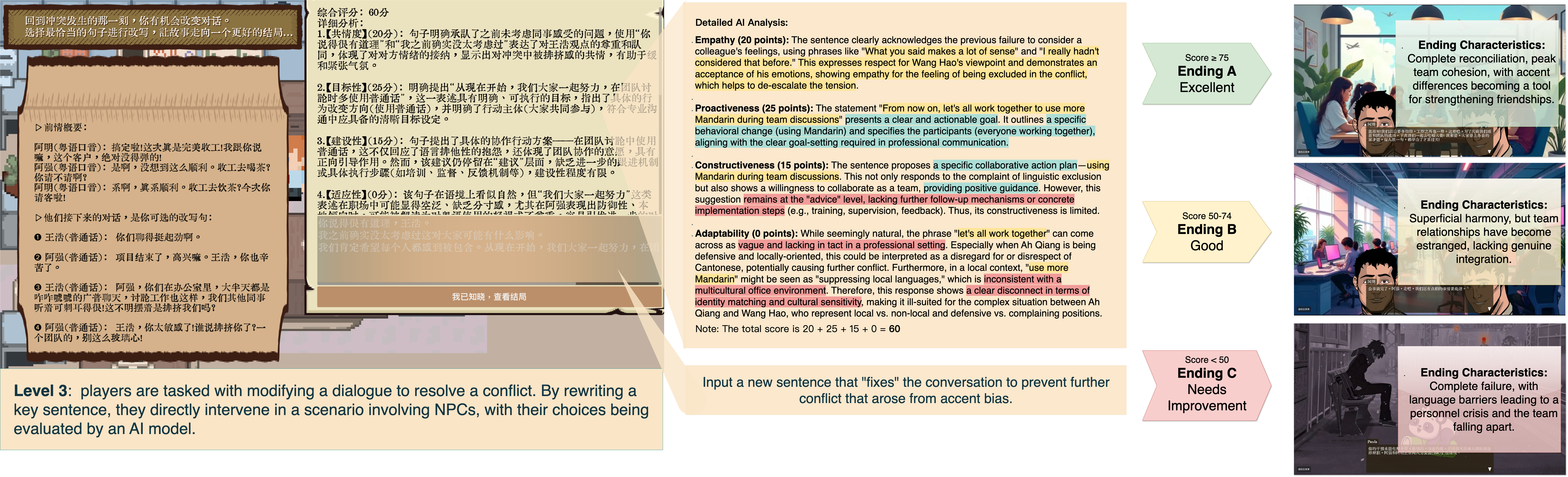}
  \caption{A branching narrative in Level 3, showcasing three possible different endings.}
  \label{fig:L3}
  \Description{This image is a sample screenshot from level 3 of a game, which has been translated from Chinese to English. In this level, players must modify a dialogue to avoid negative consequences resulting from accent bias. The image shows a detailed, AI-generated evaluation of the player's modified dialogue, with scores from four aspects: empathy, proactiveness, constructiveness, and adaptability. Each category is worth 25 points, with a total possible score of 100. The AI's evaluation is as follows: 1) Empathy (20 points): The player's sentence, which includes phrases like "What you said makes a lot of sense" and "I really hadn't considered that before," successfully acknowledges the colleague's feelings and shows respect for Wang Hao’s perspective. By accepting his emotions and demonstrating empathy for his feeling of being excluded, the dialogue helps to de-escalate the tension. 2) Proactiveness (25 points): The statement "From now on, let's all work together to use more Mandarin during team discussions" is a clear and actionable goal. It outlines a specific behavioral change (using Mandarin) and specifies the participants (everyone working together), which aligns with the clear goal-setting required in professional communication. 3) Constructiveness (15 points): The sentence proposes a specific, collaborative action plan—using Mandarin during team discussions. This not only responds to the complaint of linguistic exclusion but also shows a willingness to collaborate as a team, providing positive guidance. However, this suggestion remains at the "advice" level, lacking further follow-up mechanisms or concrete implementation steps (e.g., training, supervision, feedback). Therefore, its constructiveness is limited. 4) Adaptability (0 points): While "let's all work together" may seem natural, it can come across as vague and tactless in a professional setting. Since Ah Qiang is being defensive and locally-oriented, this could be interpreted as a disregard for or disrespect of Cantonese, potentially causing further conflict. Furthermore, in a local context, "use more Mandarin" might be seen as "suppressing local languages," which is inconsistent with a multicultural office environment. The response therefore shows a clear disconnect in cultural sensitivity, making it ill-suited for the complex situation between Ah Qiang and Wang Hao, who represent opposing local vs. non-local and defensive vs. complaining positions. With a total score of 60, the AI not only provides positive validation and suggestions for improvement but also references the player's input text multiple times to personalize the gaming experience. The image also showcases the branching-scenario mechanism. Players with a score of 75 or above reach ending A, which represents "Excellence" (complete reconciliation and peak team cohesion). Players with a score between 45 and 75 reach ending B, which represents "Good" (superficial harmony but estranged team relationships). Finally, players scoring below 45 reach ending C, which represents "Needs Improvement" (complete failure, with language barriers leading to a personnel crisis and the team falling apart). Each scene shown in the figure illustrates the different consequences of the player’s dialogue modification.}
\end{figure*}

\paragraph{Gameplay in Each Case}
The core gameplay in level 3 is a \textbf{branching narrative}, which is also the most challenging part. In each round, players must choose and correct a key sentence in the dialogue. This dialogue-rewriting mechanic functions as a form of "procedural rhetoric", persuading players to understand communication rules through interaction and constraint \cite{Bogost2007}. To ensure rigorous and objective evaluation, we used an internal algorithm based on a \textbf{Large Language Model} to provide real-time scores for player input. Unlike traditional keyword-matching systems, this algorithm uses a few-shot learning approach to conduct a multi-dimensional, deep semantic analysis of the player's rewritten sentence. This results in a composite score that determines the narrative's branching path. This process aims to achieve the game's core goal: \textbf{to cultivate players' practical skills for conflict resolution, along with creative problem-solving and communication abilities.} Ultimately, the three distinct endings—\textbf{Excellent, Good, and Needs Improvement}—correspond to the players' formation of behavioral intentions and the development of self-efficacy in addressing bias-related conflicts. An example of gameplay at this level is shown in Figure \ref{fig:L3}.

\subsection{Implementation}
\begin{figure*}
    \centering
    \includegraphics[width=\linewidth]{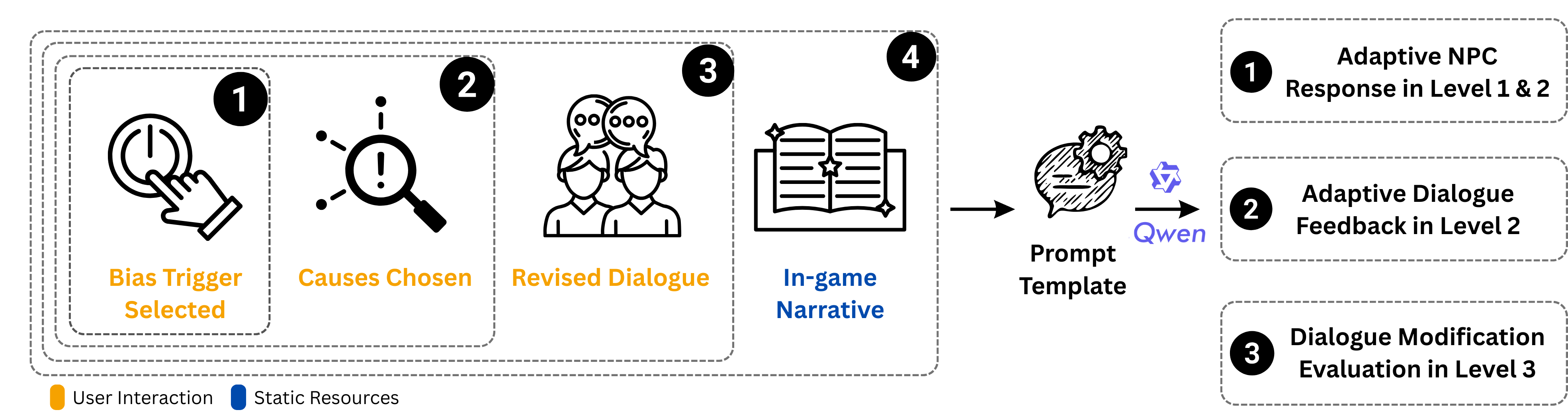}
    \caption{Overview of Local AI Interactions in CompassioMate (see supplementary for predefined prompt templates).}
    \Description{This diagram illustrates the process flow for generating dynamic content, dialogue and feedback, within the serious game CompassioMate using LLMs. The process involves four sequential steps within the game environment: 1) Bias Trigger selected (a user interaction, likely a player choosing a sensitive scenario or dialogue option); 2) Causes Selected (static resources or pre-defined elements that define the context of the bias); 3) Dialogue Modified (the actual dialogue content, potentially reflecting the player's attempted intervention); and 4) NPC's Story (static background story resources providing context for the Non-Player Character). These four inputs are consolidated into a Prompt Template, which is then fed into the LLM (represented by the gear-head icon labeled Qwen) to generate three distinct outputs for different analytical levels: 1) Adaptive NPC Response in Level 1 & 2 (likely the immediate reaction of the NPC to the player's choice in the game); 2) Adaptive Dialogue Feedback in Level 2 (providing guidance or critique on the player's chosen dialogue); and 3) Dialogue Modification Evaluation in Level 3 (a more formal assessment of the player's intervention strategy).}
    \label{fig:llm}
\end{figure*}

CompassioMate was developed using the Godot engine. Furthermore, we integrated Qwen-3-4B API into CompassioMate to power our three core conversational mechanics: adaptive NPC responses, adaptive dialogue feedback generation, and the final dialogue modification evaluation. Figure \ref{fig:llm} illustrates this architecture, showing how the system combines various static resources and real-time user inputs with predefined prompt templates to drive the LLM's outputs.
\section{Evaluation Methods of CompassioMate}
We assessed CompassioMate to meet our third research objective by conducting a three-week field study with 20 participants, with ethical approval from the first author’s institution. The study’s methodology is structured to first detail participant recruitment, followed by an explanation of the procedures and qualitative data collection. Building directly on the necessity for psychological safety and the research objective of skill development, our study investigates the following Research Questions:

\begin{itemize}
    \item \textbf{RQ1:} How does CompassioMate enable players to engage in indirect intergroup contact and foster more nuanced perceptions of individuals with different accents?
    \item \textbf{RQ2:} In what ways does interacting with CompassioMate influence players' understanding and reflection on compassionate behavior?
    \item \textbf{RQ3:} What design elements of CompassioMate are most effective or could be improved to better support empathy and perspective-taking?
\end{itemize}

\subsection{Participants}
Participant recruitment was conducted through a public advertisement posted on the social media platforms of a university in China. A total of 20 participants were recruited (P1 - P20; 8 males, 12 females). All participants were university students aged 18–24. Prior to the evaluation, demographic information was collected from participants (see Table 1). Each participant who completed the study received a ¥50 coffee voucher.  Consistent with the strategic rationale established in the Formative Study, we continued to recruit university students (18–24) as our primary participant group. This choice recognizes their critical role as a pivotal social cohort whose current language practices influence future societal trends \cite{Jiang2018}.

\begin{table}
  \centering
  \caption{Participant Demographics}

  \label{tab:demographics}
  \begin{tabularx}{0.45\textwidth}{lX} 

    \toprule

    \textbf{Characteristic} & \textbf{Count / Description} \\

    \midrule

    Total Participants & 20 \\

    Gender & Male (40\%), Female (60\%) \\

    Age Range & 18-24 \\

    Education Level & Undergraduate (95\%), Postgraduate (5\%) \\

    Major & \begin{tabular}[t]{@{}l@{}}Engineering (25\%),  Natural Sciences (10\%), \\Social Sciences (15\%),  Business (25\%),\\ Computer Science (20\%), Humanities (5\%)\end{tabular} \\

    Native Dialect & \begin{tabular}[t]{@{}l@{}}Wu (30\%), Yue (15\%), Xiang (15\%), \\ Mandarin (20\%), Gan (10\%), \\Min (5\%), Hakka (5\%)\end{tabular} \\

    \bottomrule
    
  \end{tabularx}

\end{table}
\subsection{Procedures and Data Collection}
Our evaluation consisted of three stages, ultimately collecting qualitative data. Notably, we organized semi-structured interviews to investigate the RQs and assess participants' attitude shifts regarding dialect discrimination, providing supplementary indicators for user experience. To encourage participants to think aloud naturally, the entire evaluation was conducted in a quiet computer lab with comfortable seating \cite{johnson2022using}.

\textbf{Stage 1 - Preparation Phase}: Participants were introduced to the study. Subsequently, the moderator explained the think-aloud method as a gameplay mode, showing example videos on YouTube. The moderator then conducted a warm-up exercise with participants using a sample in-game scenario to ensure they correctly understood and mastered the methodology. Following this, participants were introduced to the data collection tools, informing them about which gameplay segments would be recorded.

\textbf{Stage 2 - Game Task Phase}: For the think-aloud study, participants independently played CompassioMate from Level 1 to Level 3 on a full-screen display. Data was collected using Screen Studio, which simultaneously recorded their gameplay screen, audio, and facial expressions. This synchronized data view was accessible only to researchers, ensuring the data collection tools did not affect the players' experience.

Throughout this phase, participants were actively encouraged to verbalize their thoughts, feelings, and actions as they played. The moderator's primary role was to remain silent and non-intrusive, observing, facilitating the think-aloud process, and ensuring the recording equipment functioned correctly. The moderator would only intervene with a neutral prompt if a participant fell silent for an extended period (e.g., over 15 seconds). This prompt, such as "Please keep talking," served to remind the participant to continue verbalizing their thoughts without directing their reasoning or offering substantive feedback, as the interviewer's only intervention in a Think-Aloud protocol is a neutral reminder to ``\text{Please think out loud}'' if the subject falls silent \cite{Reinhart2022ThinkAloud}.

\textbf{Stage 3 - Post-session Interview}: The moderator conducted a semi-structured interview. This stage was designed to capture participants' in-depth cognitive processes during gameplay and their learning outcomes. All interviews were conducted offline in an undisturbed environment. The moderator's key steps involved probing participants on noteworthy gaming moments from their think-aloud session, gathering overall impressions, and soliciting concrete suggestions, thereby collecting rich, detailed feedback. All interviews were audio-recorded.

\subsection{\textbf{Data Analysis}} Our data analysis is grounded in the principles of Verbal Protocol Analysis (VPA) \cite{ericsson1993protocol}, adapted for the rich, multimodal data generated by our Think-Aloud methodology. The think-aloud sessions generated approximately 18 hours of screen, audio, and facial expression recordings in total, which allowed us to analyze participants' cognitive processes and learning outcomes. We employed a multi-stage analysis to systematically investigate our research questions.

\begin{enumerate} \item \textbf{Transcription:} All audio recordings were transcribed verbatim to create a foundational dataset for linguistic analysis. \item \textbf{Verbal Protocol Analysis (VPA) Application:} We applied VPA to understand participants' thought processes by analyzing referring, assertion, and script phrases to reconstruct engagement, problem-solving strategies, and evolving understanding of bias. 
\item \textbf{Qualitative Coding and Analysis:} Following a systematic thematic analysis approach \cite{Braun2006}, we systematically analyzed interview transcripts to identify and refine themes related to RQ1 (engagement in indirect intergroup contact and foster nuanced perceptions), RQ2 (shifts in compassionate behavior understanding), and RQ3 (design feedback). The two authors engaged in collaborative coding, where both were involved in coding the entire dataset to reflexively account for their different perspectives and assumptions. Through iterative discussion, the coders performed a ``theme-off,'' where they presented candidate themes and interrogated them to triangulate their interpretations and identify the most meaningful themes that collectively provided a coherent narrative of the data. As this analysis was more complex and we wanted to report results quantitatively \cite{McDonald2019Reliability}, we calculated IRR using Cohen’s Kappa ($\kappa$) \cite{Cohen1960} for each code; The annotators achieved an overall average inter annotator agreement of 0.86 on the agreement sample, indicating an excellent level of agreement \cite{Fleiss2003} (see supplementary for the full codebook and overall average $\kappa$ calculation). The two authors then reconciled all disagreements, a  process that refined our initial coding framework by consolidating 16 initial codes into a final, robust set of 14. This finalized codebook was then applied to the full dataset, during which a small number of new codes were inductively created to capture features not previously represented.
\item \textbf{Triangulation of Multimodal Data:} We cross-validated findings from the VPA of gameplay data with the final coded interview data, using the contextual non-verbal data annotations to strengthen or refine the interpretation of emotional and cognitive themes. This rigorous triangulation ensured the robustness of our conclusions by integrating insights from real-time verbal, non-verbal, and retrospective reflections. \end{enumerate} This comprehensive approach allowed us to move from raw data to actionable insights, addressing our research objectives by understanding how CompassioMate influences players' identification of bias and development of mitigation strategies.
\section{Findings}
This analysis utilizes \textbf{Verbal Protocol Analysis (VPA)} by examining participants' referring phrases, assertion phrases, and problem-solving scripts. The transcript is also qualitatively coded to identify themes for each research question. The IRR, validated by an overall Cohen's Kappa ($\kappa$) of 0.86 across all codes, establishes the trustworthiness of our coding framework and the findings presented below. Through the triangulation of the in-game VPA data with participants' reflective statements, we developed a comprehensive framework for interpreting the cognitive and attitudinal shifts prompted by the gameplay. This framework directly informs the discussion of CompassioMate's design mechanisms for bias mitigation and empathy promotion. All player statements are attributed to the specific participant who voiced them. In this section, we present the key findings that directly address our research questions and fulfill our third research objective.
Table \ref{tab:qualitative_findings} provides a summary of our key qualitative findings, which we will elaborate on in the following subsections.

\newcolumntype{Y}{>{\RaggedRight\arraybackslash}X}
\newcolumntype{Z}{>{\RaggedRight\arraybackslash}p{3.5cm}} 
\begin{table*}[htbp]
    \centering
    \caption{Summary of Key Qualitative Findings Grouped by Research Question}
    \label{tab:qualitative_findings}
    \begin{tabularx}{\textwidth}{Z | Y | Y}
        \toprule
        \textbf{Research Question} & \textbf{Core Insight} & \textbf{Verbatim Evidence} \\
        \midrule
        \textbf{RQ1: How does CompassioMate enable players to engage in indirect intergroup contact and foster more nuanced perceptions of individuals with different accents?} & Players progress from vague discomfort to the \textbf{precise analytical labeling} (R01-LABEL, R01-CAUSE) of bias and engage in \textbf{iterative refinement} (R01-REFINE) to formulate constructive intervention strategies (R01-FORM). & "It was an eye-opener how the game dissected the bias into technical terms like 'social classification' and 'fluency assumption.' It made me realize the problem is structured, not just rudeness." (\textbf{P11} / R01-CAUSE) \\
        \cmidrule{2-3}
        & Effective intervention requires moving past vague apologies to \textbf{proposing concrete, collaborative solutions} that repair trust and establish inclusive norms. & "The practice allowed me to move from being reserved to forming a clear strategy. Now, when I hear an unfamiliar accent, my first thought is to communicate more, not just jump to being suspicious." (\textbf{P14} / R01-FORM, R01-REFINE) \\
        \midrule
        \textbf{RQ2: How does the game influence reflection and understanding (e.g., self-awareness, emotional resonance)?} & The game fosters deep \textbf{vulnerability empathy} (R02-PERSP) for discriminated characters, leading to a strong sense of \textbf{responsibility to intervene} (R02-PRACT). & "I felt so bad for the character who couldn’t get their words out because of their accent." (\textbf{P12} / R02-PERSP) \\
        \cmidrule{2-3}
        & The experience triggered personal \textbf{self-auditing} (R02-SREF), with participants recalling and evaluating their own past roles as both perpetrators and recipients of microaggressions. & "I suddenly remembered telling a friend, 'Your Suzhou dialect sounds so soft.' Now I realize that was \textbf{harmful} and a form of \textbf{microaggression} that I must avoid in real life." (\textbf{P6} / R02-SREF) \\
        \midrule
        \textbf{RQ3: What design elements (game structure, AI role, content) are effective and could be improved?} & The progressive, three-tiered structure (R03-STRUCT) was effective, and the objective, non-judgmental AI mediator (R03-AIROLE) created a \textbf{safe environment} (R02-SAFE) critical for practice and efficacy. & "It feels like having a \textbf{neutral referee}... The AI isn't from any region, so its feedback is completely unbiased and really spot-on." (\textbf{P16} / R03-AIROLE) \\
        \cmidrule{2-3}
        & The \textbf{immediate, specific AI scoring} (R03-FEED) and safe space for trial-and-error significantly increased players' \textbf{self-efficacy} (R02-SAFE) in tackling real-world bias. & "I found myself being a lot bolder in the game than in real life... there's no real-world consequence." (\textbf{P8} / R03-FEED, R02-SAFE) \\
        \bottomrule
    \end{tabularx}
\end{table*}

\subsection{RQ1: Fostering Nuanced Perceptions and Indirect Intergroup Contact}

CompassioMate effectively guides players in identifying, analyzing, and mitigating accent bias through a structured, three-step process that players found to be well-designed and logical (P1). This process directly enables the goal of \textbf{indirect intergroup contact} by providing a vicarious, situated experience that fosters nuanced perceptions of dialect-related issues. Players move from initial emotional reactions to sophisticated, actionable strategies by engaging with realistic scenarios that highlight common, real-world problems. The game’s focus on situations like a person from the Northeast being rejected at a market or colleagues being misunderstood for speaking Cantonese (P13) made players more aware of issues they might have previously overlooked, achieving the effect of \textbf{extended contact} through simulation.

\subsubsection{Identification and Labeling of Bias}
Participants demonstrated a clear progression in their ability to identify and label nuanced forms of bias. First, through \textbf{Referring Phrase Analysis}, players initially described biased encounters with general emotional terms, as when P1 stated, "This kind of talk makes me uncomfortable." As they progressed, they adopted more precise, analytical language. P1 labeled a microaggression a "prejudice bomb," and P17 correctly identified an interaction as a "typical example of social classification prejudice." This linguistic shift indicates a cognitive move from simply feeling the effects of bias to acquiring the vocabulary for nuanced perceptions of its mechanics. Furthermore, through \textbf{Assertion Phrase Analysis}, players made clear declarative statements showing their identification of bias. For instance, P1 asserted of a specific sentence, "it's obviously a bit of a stereotype and prejudice", while P13 concluded, "This kind of thinking is most likely to be a prejudice against migrant speakers".

\subsubsection{Addressing Underlying Causes}
The game teaches players about the root causes of bias by providing direct feedback and creating scenarios that prompt deeper analysis. For example, after making incorrect choices and seeing explanations, players experienced "aha" moments (P2) and felt the game broadened their knowledge (P9). P4 realized that discrimination can be divided into many types, such as social categorization and language fluency. P20 summarized this shift: "The game really pushed me to dig deeper and understand not just what the bias is, but where it comes from and what to actually call it." This engagement with the structural causes represents a deeper \textbf{nuanced perception} of the intergroup conflict. Besides, participants began to connect specific accents to broader stereotypes. P13 reflected that people hold a stereotype that "if you speak softly, you are weak, and this unconsciously links accent to dialect." P20 correctly analyzed that a character from Shanghai might feel that people from "underdeveloped place[s], their quality and culture, will be defaulted by people from Shanghai to be not very high," identifying condescension as an underlying cause.

\subsubsection{Development of Mitigation Strategies}
A core component of the game is the dialogue-rewriting mechanic, which facilitates the development of concrete mitigation strategies. This feature enabled players to feel they could genuinely help ease tension (P1) and made them feel "very involved" and empathetic as they had to rephrase language to reduce harm (P4). This act of rewriting dialogue serves as a form of \textbf{practicing compassion} that constitutes the active element of \textbf{indirect contact}. First, P1's journey through the final challenge illustrates this process. He began with a simple strategy, declaring that "whatever" must be removed. When this strategy failed and led to a negative outcome, it prompted P1 to reflect deeply, stating, "we must stand in other people's shoes to consider some issues." His second attempt was far more nuanced, including expressing intimacy ("little Hao"), affirming a colleague's contribution, offering a sincere apology, and proposing a concrete future action: "Now, when I hear an unfamiliar accent, my first thought is that we should try to communicate more, not just jump to being suspicious." (P14). This iterative process (\textbf{R01-REFINE}) directly supports the discussion on AI-driven systems by demonstrating how the dynamic feedback loop leads to stable, refined, and actionable communication skills. Besides, the feedback system was cited as particularly helpful because it provided a score and breakdown that told players what they were missing (P1, P5). For example, P1 was told to add an apology or a promise to use standard Mandarin. P13 reflected that the feedback helped her realize her initial solution wasn't practical, prompting her to improve her own speaking habits. The scoring mechanism provided strong positive feedback (P9) and motivated players like P18 to revise their language to achieve a perfect score.
\subsection{RQ2: Understanding How CompassioMate Influences Players' Compassionate Understanding and Reflection}
Interacting with CompassioMate significantly influences players' understanding of and reflection on compassionate behavior by highlighting the real-life impacts of accent bias and providing a safe space to practice empathy and communication. The experience was described as "very meaningful" (P1) and encouraged reflection on how language can create communication barriers.

\subsubsection{Shift Towards Perspective-Taking and Empathy}
The game's scenarios consistently encouraged players to step into the shoes of others, fostering empathy. First, through \textbf{empathy Assertion}, P1 resonated with the feeling of exclusion, stating that "Seeing subtle bias from the victim's side really hit home for me — it's so much more harmful than I ever realized" (P9) , and later concluded "we must stand in other people's shoes to consider some issues." This demonstrates a conscious shift towards perspective-taking. Besides, in terms of \textbf{Emotional Impact}. Players accurately verbalized the potential emotional state of the characters, demonstrating a refined capacity for sociolinguistic empathy. P6 correctly intuited that being met with an ambiguous dismissal (such as being constantly told "human") would make someone feel "\textbf{very frustrated, and wronged}." This depth of understanding enabled emotional repair; one participant described feeling "truly empathetic" and happy when he successfully resolved a conflict (P18).

\subsubsection{Defining and Practicing Compassionate Communication}
The game's feedback mechanisms and dialogue modification process helped players deconstruct and practice compassionate communication. First, P1 analyzed that a better response required a more gentle and detailed tone, confirmation of the colleague's contribution, and a sincere apology. This shows he was internalizing a framework for compassionate communication. P1 emphasized the power of an apology, musing, "Wow, when this is said, I think it will resolve a lot of the knots between them" and concluding that one should "take the initiative to apologize" for improper behavior. This process prompted players to reflect on their own habits. P13 realized her initial solution was "not very practical," which led her to reflect on her daily speaking habits. P7 realized they might hold personal stereotypes about less developed areas and hoped to correct this in the future. The game allows players to "take action and intervene in situations," which was seen as great practice for real life (P8).

\subsubsection{The Role of the AI in a Safe Environment}
The game's AI played a crucial role by providing a "low-risk interaction" that made players more willing to express their true thoughts (P5, P9). First, the AI's responses were described as insightful and truly understanding (P16, P11), empathetic (P18), and rational (P18). It not only pointed out shortcomings but also highlighted strengths, which encouraged players to keep trying. Besides, the AI's role as a "\textbf{neutral referee}" (16) was effective because it provided objective opinions without any regional bias, helping players learn more effectively about prejudice and communication (P9). P1 reflected that "language is not just a tool for communication, it is also the cornerstone of trust." He continued that when language achieves "barrier-free communication and a tool for accurately transmitting one's emotions, then this group will become more and more close." The consistent coding of the AI as a "neutral referee" and the recurrent theme of the "safe environment".

\subsection{RQ3: Examining Our Design Considerations for Empathy and Perspective-Taking}
\subsubsection{Effective Design Elements}
One of our most effective design elements is the game’s progressive structure, particularly the final section where players can edit dialogue and receive detailed analysis and scoring (P1, P4, P18). This interactive element, which requires players to genuinely consider both sides of a conversation, is highly effective for promoting empathy. We also found that the use of familiar scenarios, such as a coffee shop or a workplace, made it easy for players to become immersed and feel a sense of closeness (P1, P13, P4). For example, P4 felt immersed because the game used common problems like getting along with colleagues.

\subsubsection{Areas for Improvement and Future Development}
We found that the geography-based hints in the first level were too difficult for some players, especially those who aren't geography experts (P1, P2). For future development, players suggested several features. First, adding emotionally intonated audio to NPC dialogues to make the experience more immersive and less like a test (P1, P5). They also suggested adding a point deduction system for mistakes to increase the challenge and player investment (P2). Lastly, we will consider shortening the initial audio clips and adding a wider variety to allow players to experience more accents (P4).

\section{Discussion}
The findings from our study presented three key insights into CompassioMate's educational approach: its effectiveness in supporting learning about bias, the role of specific design elements in shaping user experiences to facilitate this learning, and critical insights from users about the game's educational model. In this section, we examine the broader implications of these findings for the HCI community, reflect on the challenges observed, and acknowledge the limitations of our work.

\subsection{Balancing User Safety with Critical Reflection on Sensitive Topics}
Effectively teaching sensitive topics like prejudice without causing psychological discomfort is a key design challenge in HCI. Our approach uses CompassioMate, an immersive serious game, to create a psychologically safe environment for this purpose. The game's narrative is set in realistic scenarios like workplace coffee shops and job interviews that are relatable to our audience. This design is supported by previous work indicating that familiar contexts can reduce player defensiveness and foster a sense of closeness \cite{zhang2025walk, ju2025toward}. By framing the learning in a low-stakes setting, we successfully enabled players to engage with the topic safely, allowing for genuine reflection instead of an emotional reaction.

We reflect that the AI-driven system further enhanced this psychological safety by acting as a "neutral referee" (P9), providing objective and non-judgmental feedback. We engineered the AI to provide feedback based on a pre-defined framework of social, cognitive, and social-psychological mechanisms of bias, echoing research on structured, bias-resistant assessment tools \cite{LevashinaHartwellMorgesonCampion2014}. This approach contrasts sharply with human-facilitated discussions, where participants might feel scrutinized or judged, making them hesitant to express their true thoughts. Our AI offered a low-risk interaction (P5, P9), which made players more willing to express their true feelings and engage in self-reflection. This balancing of user safety with critical reflection is a key innovation, as it enabled players to transition from a passive, emotional reaction to a sophisticated, analytical understanding of bias (P1, P10). We believe that leveraging AI in this way can create a conducive environment for self-reflection and learning that is difficult to achieve in human-facilitated discussions on topics that often trigger defensive responses.

\subsection{Applying AI-Driven Dialogue Systems with Serious-Game-Based Educational Approaches for Promoting Empathy}
Our core contribution lies in demonstrating how a generative AI-driven dialogue system can be innovatively integrated into serious games to move beyond the limitations of static, rule-based feedback, thereby extending prior work on listener-focused training \cite{RovettiSumantryRusso2023, BradlowBent2008, BoduchGrabkaLevAri2021}. The system achieves this by leveraging the AI to create a dynamic, iterative, and personalized feedback loop situated within complex social bias scenarios, utilizing a dialogue-rewriting mechanic as a core component of the game. It builds on prior work that utilizes LLMs for persuasive messages \cite{wu2024mindshift} or coaching \cite{joerke2025gptcoach}, but applies these mechanisms to situated, social bias scenarios. This active virtual perspective-taking method is supported by evidence that such interventions can lead to a more stable and lasting attitude change compared to traditional methods \cite{Herrera2018LongTermEmpathy}. Consequently, the iterative and low-stakes practice addresses a key gap in existing research by creating a socially situated space where users can experiment with different responses without fear of judgment \cite{zhang2025walk}.
Furthermore, the AI's role in providing detailed and empathetic feedback was critical. Participants described the AI’s responses as "human-like" (P7), "empathetic" (P18), and "rational" (P18)—characteristics that facilitated moving beyond simple trial-and-error. This empathetic and non-judgemental feedback loop allowed players to internalize a structural understanding of compassionate communication (P1). This marks a departure from static, reactive educational games by creating a learning environment that not only documents bias but proactively teaches complex social skills.
\subsection{Challenges in the Practical Application of CompassioMate}
A significant challenge in developing an "edutainment" game like CompassioMate lies in balancing engaging gameplay with effective educational content. The perceived passivity of the initial prototype's reflection segment (which some players felt was “slightly lacking”) highlights a persistent design tension in serious games between immersion and structured pedagogical review.

We found that our use of AI, while central to the game's design, presented its own set of practical challenges. The initial prototype's "preset" answers lacked substantial impact on the narrative. To address this, we achieved a new design goal of using “generative dialogue” in a dynamic system. This approach provides constructive feedback in a "fluid, non-prescriptive way," directly addressing a gap in prior static, rule-based systems. It fosters a greater sense of agency and makes the learning process more personal and meaningful.

Another concern within the adopted framework of Intergroup Contact Theory is that, although contact has shown to be effective for more prejudiced individuals \cite{Allport1954}, the gaming experience may produce opposite effects on some players. Dixon et al. (2007) warn that contact can paradoxically weaken marginalized groups' motivation to engage in collective action against systemic inequality \cite{Dixon2007CollectiveAction}. Within our game, while the results indicate that most players reported safe contact experience, the discriminatory scenarios required for narrative realism risk demotivating some marginalized players. As P17 noted,  ``I just knew we Henanese always end up being viewed as the backward, rural farmers... These stereotypes have existed for as long as I can remember... Why bother changing minds that have never seen us differently?''.

Finally, while our game was effective in increasing awareness of accent bias, participants did report a "perspective change" and that the game prompted a "shift in attitudes toward accents," noting that it successfully helped them "uncover hidden prejudices.", we did not evaluate whether these changes extended beyond the game environment. We discuss this limitation regarding the lack of long-term behavioral evidence in detail in Section 8.4.

\subsection{Limitations and Future Work}

First, we recognize the methodological constraints of our qualitative design. Specifically, the absence of a quantitative pre-intervention assessment using validated instruments means we cannot conclusively determine the initial level of prejudice held by participants, nor can we provide a statistical measure of attitude change.

Regarding internal validity, while we acknowledge the influence of demand characteristics \cite{Orne1962OnTS}, participants were informed they were evaluating a "Geoguessr-style game" for usability, thereby avoiding explicit mention of bias change as the primary study goal \cite{Weber1972Subject}. Furthermore, while we acknowledge the potential for Hawthorne effects arising from participants' awareness of observation \cite{Harrell2013Mitigating}, our moderators adopted a silent and non-intrusive presence during sessions to mitigate this influence.

In terms of scope, this study was limited to evaluating the immediate effects of the intervention. That is, while participants reported an initial attitude shift, we did not collect empirical data on sustained behavioral change. However, most bias reduction effects are highly transient and decay rapidly within days \cite{Lai2014ReducingBias}. Nevertheless, the positive initial outcomes observed in this study suggest the potential of the AI-mediated serious game environment as a form of Indirect Intergroup Contact to mitigate bias. Consequently, a critical avenue for future research is to investigate the long-term stability of such changes and the transfer of learning from simulated environments to real-world social behavior \cite{WoutersVanNimwegenVanOostendorpVanDerSpek2013}.

Finally, the participant pool consisted exclusively of college students. While this demographic is valuable for initial design validation and understanding emerging professional biases, this homogeneity limits the generalizability of our findings. More specifically, the results may not reflect the severe social and forensic impacts of accent discrimination experienced by primary victims (e.g., rural migrants, working-class speakers) \cite{Paver2025AccentJudgements}. Furthermore, we also acknowledge self-selection bias, as individuals who volunteer for bias-reduction studies may differ systematically from the general population \cite{Zhu2022}, which further constrains generalizability. Additionally, the study’s cultural specificity poses a limitation; our scenarios are embedded in Chinese dialect politics and social contexts, which may limit the direct transferability of the findings to Western or other cultural domains where linguistic bias operates differently.

\section{Conclusion}

In this study, we examined how CompassioMate, an AI-mediated dialect-aware serious game, helps learners notice, analyze, and repair everyday bias in a psychologically safe space. Through a qualitative study, participants progressed from vague discomfort to precise bias labeling and practiced dialogue revisions that de-escalated harm, growing more confident in compassionate intervention. Empathetic guidance and transparent explanations fostered reflection, while consequence-aware branching made the impact of wording concrete. Overall, these results suggest a structured, learnable process from awareness to action. Framed by ICT, our work demonstrates how structured, positive intergroup contact can be operationalized in a digital intervention. Our study contributes to HCI by articulating a reusable design pattern for dialect-aware empathy games and distilling implications for trustworthy AI mediators. These can inform future research on psychologically safe facilitation around sensitive topics and on shaping generative, personalized feedback loops within serious games.


\begin{acks}
This work was supported by the Xi'an Jiaotong-Liverpool University Research Development Fund~RDF-22-01-062.
\end{acks}

\bibliographystyle{ACM-Reference-Format}
\bibliography{references}









\end{document}